\documentclass[prb,twocolumn,showpacs,groupedaddress,amsmath,amssymb]{revtex4}
\usepackage{ulem}
\usepackage{graphicx}
\usepackage{dcolumn}
\usepackage{bm}

\renewcommand{\k}{\mathbf{k}}
\newcommand{\E}{\mathbf{E}}
\newcommand{\HH}{\mathbf{H}}
\newcommand{\G}{\mathbf{G}}
\renewcommand{\r}{\mathbf{r}}
\newcommand{\g}{\mathbf{g}}

\begin{document}
\normalem

\title{Electromagnetic modes of a disordered photonic crystal}

\author{Vincenzo Savona}
\email{vincenzo.savona@epfl.ch}
\affiliation{Institute of Theoretical Physics, Ecole Polytechnique F\'ed\'erale de Lausanne EPFL, CH-1015 Lausanne, Switzerland}

\date{\today}

\begin{abstract}
We present a systematic numerical approach to compute the eigenmodes and the related eigenfrequencies of a disordered photonic crystal, characterized by small fluctuations of the otherwise periodic dielectric profile. The field eigenmodes are expanded on the basis of Bloch modes of the corresponding periodic structure, and the resulting eigenvalue problem is diagonalized on a truncated basis including a finite number of Bloch bands. The Bloch-mode expansion is very effective for modeling modes of very extended disordered structures in a given frequency range, as only spectrally close bands must be included in the basis set. The convergence can be easily verified by repeating the diagonalization for an increased band set. As illustrations, we apply the method to the W1 line-defect waveguide and to the L3 coupled-cavity waveguide, both based on a photonic crystal slab with a triangular lattice of circular holes. We compute and characterize the eigenfrequencies, spatial field profiles and radiation loss rates of the localized modes induced by disorder. For the W1 waveguide, we demonstrate that radiation losses, at the bottom of the spatially-even guided band, are determined by a small hybridization with Bloch modes of the spatially-odd band, induced by disorder in spite of their frequency separation. The Bloch-mode expansion method has a very broad range of applications: it can be also used to accurately compute the modes of structures with systematic modulations of the periodic dielectric constant, as in several designs of high-Q cavities.
\end{abstract}

\maketitle

\section{Introduction}

Thanks to the periodic spatial dependence of their dielectric constant, photonic crystal (PHC) structures\cite{Joannopoulos2008,John1987,Yablonovitch1987} make it possible to engineer the confinement and propagation of the electromagnetic field, and hold great promise for applications to integrated optics.

Similarly to atomic crystals, PHCs always present some amount of disorder as a result of the fabrication process.\cite{Skorobogatiy2005} When considering PHC slabs -- a periodic array of cylindric holes in a dielectric slab waveguide -- disorder arises from the irregular shape of the holes and from the roughness of the dielectric interfaces. Disorder is mostly considered as a disturbance, in the context of applications. It prevents ballistic light propagation, by inducing losses due to scattering,\cite{Engelen2008,Hughes2005,Kuramochi2005,LeThomas2008,Mazoyer2009,Mazoyer2010,O'Faolain2007,Patterson2009,Povinelli2006,Povinelli2004,Tanaka2004,Wang2008} and possibly strong localization of light,\cite{Engelen2008,LeThomas2009a,Mookherjea2008,Sapienza2010,Schwartz2007,Topolancik2007,Vlasov1999,Wang2008,Bertolotti2005,Sebbah2006} thus constituting a major physical limit to the use of PHCs as interconnects in photonic circuits, and in all applications relying on slow light propagation.\cite{Krauss2007,SoljaCiC2004,Baba2008} Disorder-induced scattering is also responsible for extrinsic radiation losses,\cite{Andreani2006,Gerace2004,Hughes2005} as it mixes Bloch momenta of truly guided and leaky modes of the nominally periodic structure. 

More generally, it has been suggested that PHCs might be an election system for studying the fundamental properties of light scattering in disordered media. It is rather singular that the seminal work by Sayeed John,\cite{John1987} considered as one of the first works to introduce the idea of a PHC, actually considers the photonic bands only as a tool to achieve Anderson localization of light.\cite{Anderson1985} The idea put forward in the work by John is that, in the vicinity of a point of high symmetry, the dispersion of a photonic band can be approximated by a parabola and Maxwell equations can consequently be mapped onto an effective Schr\"odinger equation in a continuous effective medium, including a random potential term derived from the actual disorder profile of the system. 

Modeling the effects of disorder in PHCs is a current research topic and a very challenging computational task. In a regular structure, Bloch's theorem allows restricting the solution of Maxwell equations to a single elementary cell. Taking advantage of the crystal periodicity, several computational approaches have been developed, based on mode expansion\cite{Andreani2007,Andreani2006,Johnson2001} or scattering matrix\cite{Botten2004,Gralak2002,Tikhodeev2002,Whittaker1999} formalisms. Disorder however breaks the translational symmetry of the crystal, and Bloch's theorem can no longer be applied. Then, a full solution of Maxwell equations over an extended region of space, possibly including thousands of elementary cells would be needed both for studying fundamental properties, such as Anderson localization, and for modeling device operation. First principle calculations based on a finite-element sampling of real space are thus of very limited use, as the computational cost of simulating such extended samples would be prohibitive. Theoretical approaches to disordered PHCs have been proposed, based on various levels of approximation. Some properties, such as diffusive light propagation and the corresponding radiation losses, can be predicted already within a perturbation theory of light scattering.\cite{Johnson2005,Gerace2004,Hughes2005,Wang2008} More generally, a variety of coupled-mode approaches have been proposed for studying disordered PHCs.\cite{LeThomas2008,Olivier2003,Patterson2009,Povinelli2004} A coupled-mode theory in general accounts for light scattering between few selected modes. The reduced number of modes makes it possible to model the scattering processes in a fully non-perturbative fashion, but restricts the predictive power of the method. As an example, Anderson localization of light arises from multiple coherent scattering between many different propagating modes,\cite{Anderson1985,Kramer1993,Vlasov1999} and typically requires a solution of Maxwell equations beyond few-mode coupling, to be modeled correctly. It should be noted however that the idea of mode coupling can be generalized to scattering theories including a full basis of propagating modes. In this case, the theory is formally equivalent to a full solution of Maxwell equations, as is e.g. the case for the quasi-normal Bloch-mode scattering matrix approach recently developed by Lecamp et al.\cite{Lecamp2007a,Mazoyer2009}

A general aspect of most theoretical approaches\cite{Botten2004,Gralak2002,Tikhodeev2002,Whittaker1999,Lecamp2007a} is the fact that they compute directly the linear response of the periodic dielectric medium. For several reasons however, it would be beneficial to compute the actual Maxwell eigenmodes. Apart from the direct visualization of the Anderson-localized spatial profiles, knowledge of the actual eigenmodes and of their detailed Bloch-mode components can give insight into the mechanisms of slow-light propagation,\cite{Baba2008,Gersen2005,Krauss2007,LeThomas2009,Notomi2004,Vlasov2005} radiation\cite{Andreani2006,Gerace2004,Hughes2005} and disorder-induced\cite{Engelen2008,Hughes2005,Kuramochi2005,LeThomas2008,Mazoyer2009,Mazoyer2010,O'Faolain2007,Patterson2009,Povinelli2006,Povinelli2004,Tanaka2004,Wang2008} losses. When modeling systems with coupled photonic and electronic degrees of freedom, such as quantum dots embedded in cavities or guides,\cite{Chauvin2009,Dewhurst2010,Lecamp2007,Lund-Hansen2008,MangaRao2007,Rao2007,Viasnoff-Schwoob2005,Sapienza2010,Gallo2008,Thyrrestrup2010} or PHC polaritons,\cite{Andreani2007,Bajoni2009} it is most natural to start with the eigenstates of both coupled subsystems, especially within a fully quantum mechanical treatment where second quantization of electromagnetic modes is needed. If the linear response function is known, then eigenmodes can be obtained by finding the poles of its analytical continuation on the complex frequency plane.\cite{Tikhodeev2002} Conversely, if Maxwell eigenmodes are known, the linear response function can always be expressed as a function of these eigenmodes, by using the resolvent representation in Fredholm theory.\cite{Tarel2010} Both indirect ways are however computationally far from being optimal. A method for directly computing Maxwell eigenmodes of an extended PHC with a small perturbation of the periodic dielectric profile would thus be an indispensable complement to linear response theories.

Here, we develop the {\em Bloch-mode expansion} formalism, for directly computing Maxwell eigenmodes of a photonic structure characterized by a small perturbation of the otherwise periodic dielectric profile. The basic idea behind this formalism is that eigenmodes of the perturbed structure can be expanded on the basis of Bloch modes of the regular, unperturbed PHC. This expansion extends, in general, to all bands and all Bloch momenta of the regular structure. If the perturbation is small, as in the case of disorder in state-of-the-art PHC slabs, then the expansion can be truncated on a relatively small subset of Bloch bands -- spectrally close to the frequency region of interest -- and the Maxwell equation can be diagonalized restricting to this subspace, at a considerably reduced computational cost. The convergence of the method can be easily assessed by repeating the calculation on a progressively augmented Bloch-band subset. After laying down the general formalism, as examples of application we study the W1 line-defect waveguide\cite{Benisty1996,Meade1994} and the L3 coupled-cavity waveguide,\cite{Jagerska2009} both based on a PHC slab structure with a triangular lattice of circular holes. For computing the Bloch modes of the regular PHC, we use the guided-mode expansion method recently proposed by Andreani and Gerace,\cite{Andreani2006} because it offers a good 
balance between predictivity and ease of implementation, although some limitations in its 
accuracy to predict out-of-plane losses have been recently put forth.\cite{Mazoyer2009} This is not however the only possible choice, and the Bloch-mode expansion could equally rely on a first-principle numerical solution of Maxwell equations for computing Bloch modes of the regular structure. We show how a fully converged calculation of modes in the guided region of both waveguides can be achieved at a minimal computational cost, for structures as long as 1024 elementary PHC cells. We study the localization of the modes and their radiation loss rates, and discuss the influence of disorder  and losses on slow-light propagation, in the light of the existing literature. For the W1 waveguide, we show that radiation losses at the bottom edge of the spatially-even guided band are determined by hybridization with the spatially-odd guided band, in spite of their mutual frequency separation. This result explains the physical limitations to the Q-factor of W1-based structures, such as ``gentle confinement'' cavities,\cite{Song2005,Felici2010,Kuramochi2006} and shows the way to a significant optimization of these systems.

In Section \ref{theory}, we present the theoretical formalism. We also discuss the method for computing radiation loss rates, when Bloch modes are computed via guided-mode expansion. In Sections \ref{application} and \ref{L3}, we discuss the results of the application of the present method to a W1 line-defect waveguide and to a L3 coupled-cavity waveguide, respectively. Section \ref{conclusions} presents the main conclusions and an outlook of future applications.

\section{Theory}
\label{theory}

We consider a PHC made of a periodic array of cylindrical air holes etched in a dielectric slab of thickness $d$. The regular PHC -- namely in the absence of disorder -- is described by the dielectric constant $\epsilon(\r)$. In presence of disorder instead, the dielectric constant is defined as
\begin{equation}
	\epsilon^\prime(\r)=\epsilon(\r)+\delta\epsilon(\r)
	\label{epsilondis}
\end{equation}
As the holes are present only in the slab, we assume that both $\epsilon(\r)$ and $\epsilon^\prime(\r)$ take constant values in the upper and lower semi-infinite claddings. In the following, we will always consider an ``air bridge'' configuration, for which $\epsilon=\epsilon^\prime=1$ and $\delta\epsilon=0$ in the claddings. Here, we are neglecting the roughness that might characterize the slab interfaces, as we are primarily interested in the disorder affecting the cylindrical holes.

The starting point of our analysis are the Bloch modes $\E_{\k n}(\r)$ of the regular PHC structure under study. They are the eigenmodes of the Maxwell equation
\begin{equation}
\nabla\times\nabla\times\E_{\k n}(\r)-\frac{\omega_{\k n}^2}{c^2}\epsilon(\r)\E_{\k n}(\r)=0\,,
\label{maxwell}
\end{equation}
where $\omega_{\k n}$ are the corresponding eigenfrequencies. These modes might have been obtained by any computational method that takes advantage of Bloch's theorem. For example, they could be computed by exact finite-element solution of Maxwell equations, or by an approximate scheme like e.g. guided-mode expansion.\cite{Andreani2006} In the examples of application that follow, we will consider a finite system size with periodic boundary conditions, thus implying a discrete set of Bloch-momentum values ${\bf k}$. The formalism presented in this Section can however be straightforwardly generalized to a continuous ${\bf k}$-spectrum.

In presence of disorder, the actual eigenmodes of the system are denoted by $\E_\beta(\r)$, and are indexed by a global index $\beta$, as the Bloch momentum $\k$ is no longer a conserved quantity. These modes are solution of the Maxwell equation
\begin{equation}
\nabla\times\nabla\times\E_\beta(\r)-\frac{\omega_\beta^2}{c^2}\epsilon^\prime(\r)\E_\beta(\r)=0\,.
\label{maxwelldis}
\end{equation}
They can be formally expanded on the basis of the Bloch modes of the regular structure:
\begin{equation}
\E_\beta(\r)=\sum_{\k n}U_\beta(\k, n)\E_{\k n}(\r)\,.
\label{expansion}
\end{equation}
By replacing (\ref{expansion}) into (\ref{maxwelldis}), and using (\ref{epsilondis}) and (\ref{maxwell}), we obtain
\begin{equation}
\sum_{\k n}U_\beta(\k, n)\left[\frac{\omega_{\k n}^2-\omega_\beta^2}{c^2}\epsilon(\r)-\frac{\omega_\beta^2}{c^2}\delta\epsilon(\r)\right]\E_{\k n}(\r)=0\,.
\label{eqU1}
\end{equation}
We take the scalar product of Eq. (\ref{eqU1}) by $\E^*_{\k^\prime n^\prime}(\r)$ and integrate on $\r$ (we assume the modes $\E_{\k n}(\r)$ to be normalized over the total volume of the system). By taking advantage of the orthogonality relation\cite{Sakoda2001}
\begin{equation}
\int d\r\epsilon(\r)\E^*_{\k^\prime n^\prime}(\r)\cdot\E_{\k n}(\r)=\delta_{\k\k^\prime}\delta_{n n^\prime}\,,
\label{orthogonality}
\end{equation}
we finally obtain
\begin{equation}
\sum_{\k^\prime n^\prime}\left[\frac{\omega_{\k n}^2-\omega_\beta^2}{c^2}\delta_{\k\k^\prime}\delta_{n n^\prime}-\frac{\omega_\beta^2}{c^2}V_{\k n,\k^\prime n^\prime}\right]U_\beta(\k^\prime, n^\prime)=0\,.
\label{eqU2}
\end{equation}
Eq. (\ref{eqU2}) is a generalized eigenvalue problem, formally equivalent to the full solution of Maxwell equation (\ref{maxwelldis}). The matrix elements $V_{\k n,\k^\prime n^\prime}$ contain all information about the disorder profile. They are defined as
\begin{equation}
V_{\k n,\k^\prime n^\prime}=\int d\r\delta\epsilon(\r)\E^*_{\k^\prime n^\prime}(\r)\cdot\E_{\k n}(\r)\,.
\label{matel}
\end{equation}
It should be pointed out that, when assuming periodic boundary conditions for the Bloch modes, the Bloch-mode expansion (\ref{expansion}) implies periodic boundary conditions also for the modes $\E_\beta(\r)$. Thus, boundary effects at the side terminations of the PHC (e.g. interface with a ridge-waveguide), that might give rise to additional spectral features in some structures,\cite{Mazoyer2010} should be computed additionally, e.g. through mode-coupling or transfer-matrix approaches that go beyond the scope of the present work.

It is interesting to recall the principle underlying the present formalism had partly been laid down already in the seminal work by S. John.\cite{John1987} There, it is pointed out that, close to a band edge, the Maxwell problem can be approximated by an effective Schr\"odinger equation whose kinetic term is essentially determined by the curvature of the single Bloch band under study. The potential term is nonlocal and is given by the matrix element of the random potential between actual electric field modes. This exactly corresponds to our Eqs. (\ref{eqU2}) and (\ref{matel}) if restricted to the edge of a single Bloch band. According to the basic theory of electronic states in crystals, restricting to the edge of a single Bloch band coincides with the effective-mass approximation. The present formalism generalizes John's original idea to an expansion over several bands. Its advantage lies in the possibility to truncate the vector space defining the eigenvalue problem (\ref{eqU2}) to a subset of the bands of the regular PHC, in order to ensure optimal convergence of the properties under study. In particular, when studying the effect of disorder on the band $n_0$, it should be possible to neglect all bands $n$ whose frequencies are much further away from that of band $n_0$, than the disorder contribution $\langle V\rangle$. By inspection of Eq. (\ref{eqU2}), it is easily shown that bands can be neglected if the condition 
\begin{equation}
|\omega_{\k_0 n_0}^2-\omega_{\k n}^2|\gg\omega_{\k n}^2|V_{\k_0 n_0,\k n}|
\label{condition}
\end{equation}
is fulfilled. The reliability of the method can be assessed by repeating the diagonalization of (\ref{eqU2}) for an increasing number of bands and checking for convergence of the eigenvalues and eigenvectors of interest. This perturbation criterion holds however only for $n\ne n_0$. Within the same band -- in particular for low-dimensional systems -- even the slightest amount of disorder can induce a strong mixing of Bloch modes in correspondence to band extrema, possibly giving rise to Anderson localization, as originally suggested by John.\cite{John1987} The typical situation in which condition (\ref{condition}) is particularly favorable is that of defect states in the bandgap of a two-dimensional PHC, for which most other bands are naturally separated by the frequency bandgap. Among the systems falling in this class are all one-dimensional systems based on a two-dimensional PHC, such as line-defect waveguides,\cite{Benisty1996,Meade1994} coupled-cavity waveguides,\cite{Stefanou1998,Yariv1999} etc. These systems are of particular interest because of the potentially high quality factor and of the possibility of slow-light propagation.\cite{Baba2008,Krauss2007} Slow-light in particular can be exploited to enhance the coupling of light to the electronic states of embedded quantum dots, for applications as single photon emitters.\cite{Chauvin2009,Dewhurst2010,Lecamp2007,Lund-Hansen2008,MangaRao2007,Rao2007,Viasnoff-Schwoob2005,Sapienza2010,Gallo2008,Thyrrestrup2010} In this context in particular, knowing the actual eigenmodes of the electromagnetic field is crucial for modeling their coupling to electronic states.\cite{Tarel2010} Otherwise, the method should also be useful to model the states of a disordered two-dimensional PHC at the edges of the bandgap.\cite{LeThomas2009a,Schwartz2007} Finally, it is also possible to apply the present method to special designs involving a small {\em systematic} perturbation $\delta\epsilon(\r)$ of a periodic PHC, as in the case of long defect cavities\cite{Felici2010} or cavities obtained by a slow grading of the lattice parameters.\cite{Song2005,Felici2010,Kuramochi2006} In the next section we will apply the present method to a W1 line-defect waveguide in a PHC made of a triangular lattice of air holes, and to the related system of a L3 coupled-cavity waveguide. In the case of the W1 waveguide, we will show that accounting for more than one Bloch band is essential in determining the properties of the Maxwell eigenmodes.

One of the main consequences of disorder in PHCs are the {\em extrinsic radiation losses}. In a regular structure, modes $\E_{\k n}(\r)$ lying outside the light cone in the $(\omega_\k,\k)$-space are truly guided within the slab and ideally lossless. Inside the light cone, on the other hand, the electromagnetic modes are characterized by a continuous frequency spectrum, with a spectral function that has narrow resonances in correspondence to the band dispersion $\omega_{\k n}$. In this case, formally the index $n$ only labels the resonance frequencies, while eigenmodes should be labelled by a continuous frequency variable. The continuous frequency spectrum is related to the lossy nature of these modes, and implies a finite rate of radiation out of the plane into the claddings. When disorder is present, Eq. (\ref{expansion}) mixes discrete and continuous parts of the spectrum, resulting in a continuous spectrum over the whole $(\omega_\k,\k)$-space. Now all modes (\ref{expansion}) have components in the continuous part of the spectrum, inside the light cone, and are therefore lossy. For these lossy modes, Eqs. (\ref{expansion})-(\ref{matel}) would not strictly hold and should be modified to account for the continuous frequency spectrum. Then, a formally exact but computationally demanding approach to evaluate the loss rates, would require solving the Maxwell eigenvalue problem (\ref{maxwell}) numerically in three dimensions, and investigating the properties of its continuous spectrum. Here we prefer to adopt a computationally less demanding approach, based on the guided mode expansion recently proposed by Andreani et al. \cite{Andreani2006} to compute the modes of the regular structure $\E_{\k n}(\r)$. These modes are expanded on a subset of the full Maxwell basis, formed by the guided modes of an effective homogeneous dielectric slab
\begin{equation}
	\E_{\k n}(\r)=\sum_{\G,\alpha}c_n(\k+\G,\alpha)\E^{(g)}_{\k+\G,\alpha}(\r)\,,
\label{gmemodes}
\end{equation}
where $\alpha$ runs over the different guided modes. The leaky modes of the effective slab are thus neglected in this expansion. This approximation turns out to be highly predictive in typical situations, at a marginal computational cost. Radiation losses of Bloch modes can then be evaluated by a perturbative approach analogous to the Fermi golde rule in quantum mechanics, involving overlap integrals between modes (\ref{gmemodes}) and leaky modes of the effective slab. The continuous spectrum of leaky Bloch modes is thus accounted for by an irreversible decay of the approximately guided modes (\ref{gmemodes}) into the leaky modes of the homogeneous slab. The same approach can be applied in the present context to compute the radiation loss rates of the eigenmodes $\E_\beta(\r)$ of the disordered structure. Then, to the real eigenvalues $\omega^2_\beta/c^2$ is associated an imaginary part $\Gamma_\beta$ related to the loss rate. The expression for this quantity is
\begin{equation}
\Gamma_\beta=\pi\sum_{\k,\G}\sum_{\lambda {\rm =TE,TM}}\sum_{j=1,3}\left|{\cal H}_{\beta\g}^{rad}\right|^2\rho_j\left(\g;\frac{\omega_\beta^2}{c^2}\right)\,,
\label{losses}
\end{equation}
where $\g=\k+\G$, $\rho_j$ is the density of leaky modes for a given frequency and in-plane momentum, and ${\cal H}_{\beta\g}^{rad}$ is the overlap integral between mode $\E_\beta(\r)$ and the leaky mode of the effective slab with momentum $\g$. The index $j$ runs over the two cladding layers, while the index $\lambda$ runs over the polarization basis of the radiative modes. From here, the actual loss rate can be computed as
\begin{equation}
\gamma_\beta\simeq c^2\Gamma_\beta/(2\omega_\beta)\,.
\label{lossrates}
\end{equation}
The explicit expressions for the quantities entering Eq. (\ref{losses}) are presented in the Appendix. One limit of validity of this perturbative approach is of course that $\gamma_\beta\ll\omega_\beta$. 

In presence of disorder, an additional condition must be fulfilled. While in a regular structure light would propagate ballistically in each Bloch mode $\E_{\k n}$, disorder is generally responsible for damping of the transmission along the plane, induced by multiple scattering between different Bloch momenta and in particular coherent backscattering.\cite{Mazoyer2009,Patterson2009} This damping is typically described as a disorder-induced loss mechanism. The same multiple scattering can induce Anderson localization of modes $\E_\beta(\r)$, and the average attenuation length of the transmission at a given frequency is a measure of the average localization length of modes at that frequency. In this picture, a competition between radiation losses and multiple scattering arises. The destructive interference in multiple scattering, at the origin of forward attenuation and Anderson localization, requires that radiation losses are negligible over the total multiple scattering path. As we plan to compute modes $\E_\beta(\r)$ based on a guided mode expansion of Bloch modes $\E_{\k n}(\r)$ -- thus without accounting for radiation losses -- then an additional condition of validity of the present approach is that the attenuation length corresponding to $\gamma_\beta$ be much longer than the localization length of modes $\E_\beta(\r)$. If this is not the case, then Eq. (\ref{expansion}) would be inconsistent with the radiation loss rate $\gamma_\beta$ computed perturbatively afterwards, as the Bloch waves involved in the expansion would be damped before producing the interference pattern that gives rise to the localized mode. This condition holds particularly well for modes that in the regular structure would lie outside the light cone and thus be radiation lossless. A typical example are the guided modes of a W1 line-defect waveguide, as discussed in the next Section.

Before proceeding to applications, we point out that the present method is fully non-perturbative. If the Bloch-mode basis, on which the expansion is made, is computed exactly, then the converged expansion (\ref{expansion}) coincides with the full solution of the Maxwell problem for the extended disordered structure. If, as we do in the examples below, the Bloch basis is computed via guided-mode expansion,\cite{Andreani2006} then modes (\ref{expansion}) are still close to the full solution of Maxwell equations, with the only assumption that the admixture of out-of-plane propagating modes is negligible and can be taken into account perturbatively at a later stage through Eq. (\ref{losses}).

\section{Application to a W1 line-defect waveguide}
\label{application}

As an example of application of the present method, we consider the problem of disorder in W1 line-defect waveguides. We choose this system, as it was widely investigated both theoretically and experimentally, thus representing a valid platform on which our method can be tested. The effect of disorder on one-dimensional systems is particularly dramatic, as all states are expected to be exponentially localized.\cite{Kramer1993} In the context of PHC waveguides, the role of disorder was highlighted already in early experimental studies. Two disorder-related effects are of primary importance and were extensively discussed in the literature. First, the restrictions on slow-light propagation, due to coherent multiple scattering that induces an attenuation of the transmitted intensity.\cite{Baba2008,Engelen2008,Gersen2005,Hughes2005,Kuramochi2005,Krauss2007,LeThomas2008,LeThomas2009,Mazoyer2009,Mazoyer2010,Notomi2004,O'Faolain2007,Patterson2009,Povinelli2006,Povinelli2004,Tanaka2004,Vlasov2005,Wang2008} Second, the appearance of extrinsic radiation losses.\cite{Andreani2006,Gerace2004,Hughes2005} In what follows, we characterize the eigenmodes of a W1 waveguide in presence of disorder and discuss their properties in the light of the existing studies. We show that the present method allows to actually compute the localized eigenmodes, to study their localization length and their radiation loss rates, and to characterize the disorder-induced transmission losses. In addition, the method provides new insight into the physical mechanisms governing radiation losses. In particular, we show that extrinsic radiation losses in W1 waveguides are mostly determined by hybridization of the spatially-even and spatially-odd guided Bloch modes -- a fact that has great impact on the design and optimization of high-Q optical cavities based on ``gentle confinement'' of W1-guided modes.\cite{Song2005,Felici2010,Kuramochi2006}

\begin{figure}
\includegraphics[width=0.5 \textwidth]{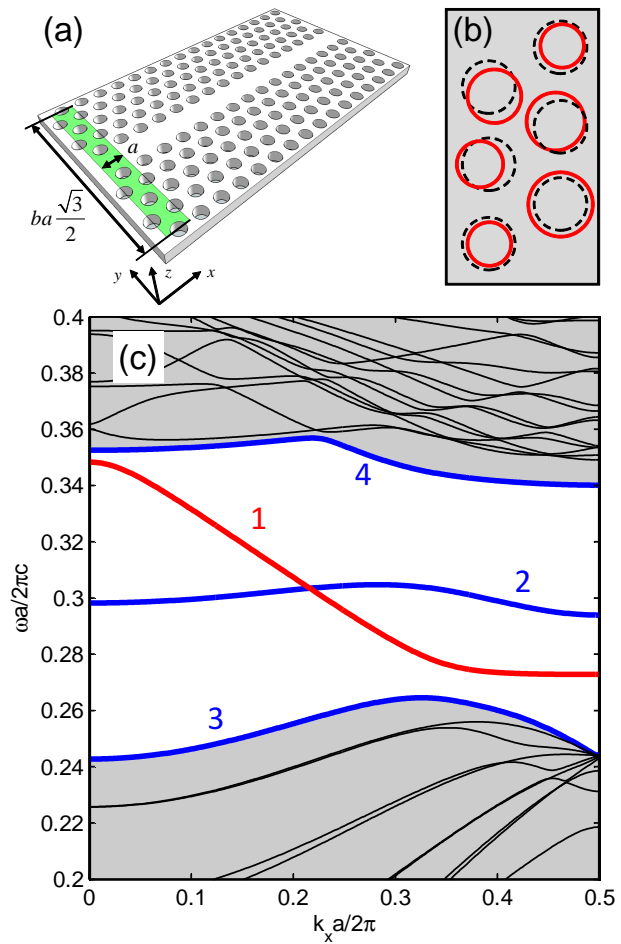}
\caption{\label{fig1} (a) Sketch of the W1 line-defect waveguide. The shaded region denotes the supercell used for computing the bands of the regular structure. (b) Sketch of the disorder model used in this work. The dashed and full circles indicate holes of the regular and disordered structure respectively. (c) Band structure of the W1 waveguide with $d=0.5a$, $\epsilon=12$, $r=0.3$, and $b=10$. Band 1 (red) is the spatially-even guided band. Bands labeled 1 to 4 are those that were included in the Bloch-mode expansion of the disordered structure.}
\end{figure}

We consider a W1 line-defect waveguide, consisting in one missing row of holes in a triangular lattice of cylindrical air holes of radius $r$, with lattice parameter $a$ and slab thickness $d$. The W1 waveguide is sketched in Fig. \ref{fig1}(a). The modes $\E_{\k n}(\r)$ of the regular structure are computed by means of guided-mode expansion.\cite{Andreani2006} The waveguide is oriented along the $x$-axis, while the $z$-axis is perpendicular to the slab. We adopt a rectangular supercell of width $a$ along ${\bf x}$ and height $ba\sqrt{3}/2$ along ${\bf y}$, where $b$ is an integer, as sketched in Fig. \ref{fig1}(a). Modes that are strongly confined into the guide core, are scarcely affected by the superperiodicity along ${\bf y}$ introduced by the supercell.\cite{Benisty1996} We can thus safely restrict to modes having $k_y=0$, corresponding to an effective one-dimensional system whose bands are characterized by Bloch momentum $k_x$ along the $x$-axis. For simplicity, in what follows we denote $k_x$ simply as $k$. For the guided-mode expansion, we restrict to the lowest guided mode of the effective slab, having TE polarization. We thus neglect the TE-TM mixing induced by the hole lattice -- an approximation that holds very well for small $d/a$, when the frequency of the next guided mode falls outside the band gap of the 2-D lattice.\cite{Andreani2006} Calculations can be easily generalized to an expansion on several guided modes if needed, by a straightforward extension of the formalism. As usual, a scaling relation with respect to the lattice parameter $a$ holds and we can express all lengths, momenta and frequencies in dimensionless units. For the calculations that follow, we set $d/a=0.5$, $r/a=0.3$, $b=10$, and we consider an air-bridge structure with dielectric constant $\epsilon_{1,3}=1$ in the cladding layers, and $\epsilon_2=12$ for the slab material, corresponding to an effective-slab  dielectric constant $\bar{\epsilon}_2=8.77$ for the calculation of the guided modes (see Appendix). In the calculations we use Ho's method \cite{Benisty1996} for the Fourier transform of the dielectric profile, and we truncate the expansion at $G_{max}=3$ (in units of $2\pi/a$), corresponding to a basis of 229 guided modes. Fig. \ref{fig1}(c) shows the band structure in the region of the guided mode. Bands labeled 1 to 4 will be included in the truncated band set used for the disorder calculations.\footnote{Throughout this work, when not differently specified, all displayed results for the W1 waveguide were obtained by including bands 1 and 2 in the calculation.} Band 1 is the main guided band, often referred to as index-guided or spatially-even band (with respect to the $\hat{\sigma}_{kz}$ mirror plane), while band 2 is the gap-guided, spatially-odd band. 

We model disorder as a gaussian fluctuation of both hole radii $r_m$ and positions $(x_m,y_m)$, where $m$ runs over all holes of the extended structure, as sketched in Fig. \ref{fig1}(b). In particular, $r_m=r+\delta r_m$, $x_m=x^{(0)}_m+\delta x_m$, and $y_m=y^{(0)}_m+\delta y_m$, where $r$ and $(x^{(0)}_m,y^{(0)}_m)$ are respectively the radius and the positions of the holes in the regular structure. The fluctuations $\delta r_m$, $\delta x_m$, and $\delta y_m$ are Gauss-distributed with standard deviations $\sigma_r$, $\sigma_x$, and $\sigma_y$. This ``minimal'' disorder model -- in particular the radii fluctuations -- is expected to catch the main effects of hole shape fluctuations on the spectral properties \cite{Skorobogatiy2005,Mazoyer2009,Mazoyer2010}, at a low numerical cost in computing the disorder matrix elements $V_{kn,k^\prime n^\prime}$, as shown in the Appendix. More refined disorder models, possibly accounting for the roughness of the hole contour \cite{Hughes2005,Gerace2005,Patterson2009,Patterson2010}, could be easily implemented. As discussed above, the converged expansion (\ref{expansion}) corresponds to the exact solution of the Maxwell problem and would thus automatically include local field effects in proximity of the hole sidewalls, whose importance has recently been debated.\cite{Andreani2007,Gerace2005,Patterson2010} In what follows, we restrict our analysis to the situation in which $\sigma_x=\sigma_y=\sigma_r=\sigma$. Although we didn't carry out a systematic study, our simulations seem to confirm that radii fluctuations are the dominant effect, while position fluctuations give a less significant contribution. We consider values of $\sigma$ ranging from $0.001a$ to $0.008a$, corresponding to state-of-the-art structures.\cite{Skorobogatiy2005,Mazoyer2010}

For the numerical simulations, we consider W1 structures of length $L=1024a$, namely including 9216 holes for the supercell height $b=10$ considered here. This corresponds to assuming a uniform grid of 1024 $k$-points in the interval $[-\pi/a,\pi/a]$ for the eigenvalue problem (\ref{eqU2}). A single disorder realization of a system of this length requires a few hours on a desktop computer. Computing Maxwell eigenmodes of such an extended structure with other first-principle methods, like e.g. a finite-element sampling of real space, would constitute a practically untractable problem. The advantage of the present approach is therefore striking.

In order to study the spectral properties of the system, a natural choice would be to introduce the quantity 
\begin{equation}
G(k,\omega) = \sum_\beta\frac{|\HH_\beta(k,k_y=0)|^2}{\omega-\omega_\beta+i\gamma_\beta}\,,
\label{prop1}
\end{equation}
where $\HH_\beta(k,k_y=0)$ is the Fourier transform, taken at $k_y=0$ (and $z=0$) of the magnetic field $\HH_\beta(\r)$ of the $\beta$-th eigenmode, and $\gamma_\beta$ is the rate of radiation losses computed according to (\ref{losses}). Eq. (\ref{prop1}) is a scalar Green's function of the electromagnetic field, corresponding to the trace of the Green's tensor of Maxwell equations.\cite{Morse1953} It is evaluated at equal initial and final Bloch momenta (forward propagation), thus providing information about the transmission amplitude and, ultimately, about the spectral density of the field.\cite{Kramer1993} It is expressed through the resolvent representation\cite{Morse1953} based on the magnetic field modes $\HH_\beta(\r)$. These modes constitute a complete orthonormal basis with the standard measure for the scalar product, as opposed to the dielectric measure $\epsilon(\r)$ needed to define a scalar product for electric field modes.\cite{Sakoda2001,Tarel2010} They are related to the corresponding electric field modes as $\HH_\beta(\r)=-i(c/\omega_\beta)\nabla\times\E_\beta(\r)$. In Eq. (\ref{prop1}) however, the choice $k_y=0$ corresponds to integrating the field along $y$, thus missing all spectral features related to spatially-odd modes with respect to the $\hat{\sigma}_{kz}$ mirror symmetry. In order to be able to visualize these spectral contributions, we prefer to adopt a different choice and introduce the quantity
\begin{equation}
G_y(k,\omega) = \sum_\beta\frac{|\HH_\beta(k,y)|^2}{\omega-\omega_\beta+i\gamma_\beta}\,,
\label{prop2}
\end{equation}
where  $\HH_\beta(k,y)$ is obtained by Fourier transforming the field $\HH_\beta(\r)$ with respect to $x$ only. Eq. (\ref{prop2}) represents the Green's function for the propagation of the field along $x$, at a fixed position $y$. In this way, by evaluating (\ref{prop2}) at $y\ne0$, i.e. displaced from the symmetry plane $\hat{\sigma}_{kz}$, spectral signatures of both even and odd modes will appear. For all plots, we choose $y=0.067a$. 

\begin{figure}
\includegraphics[width=0.5 \textwidth]{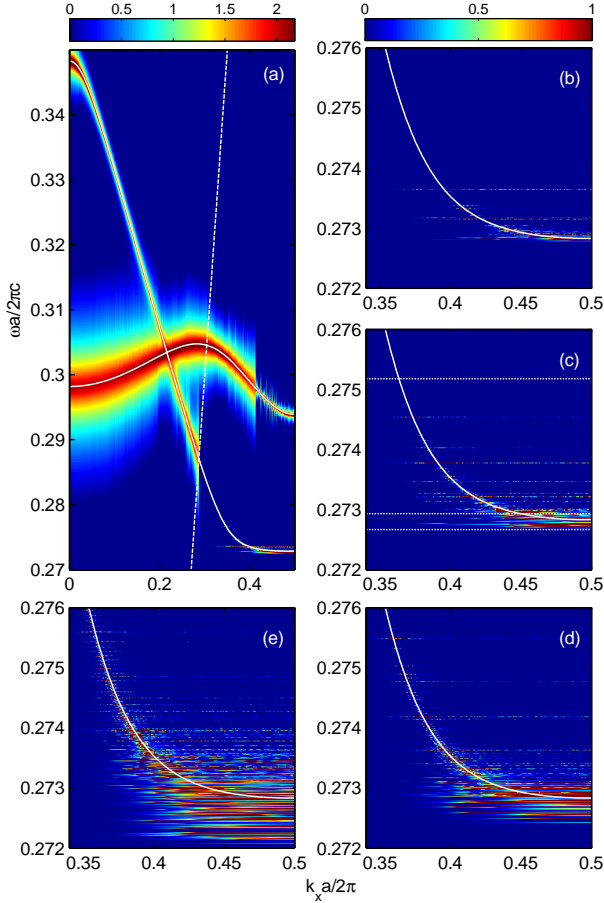}
\caption{\label{fig2} Spectral function (see text) of the disordered W1 waveguide. (a) $\sigma=0.002a$, plotted on a $\log_{10}$ color scale. Bands 1 and 2 of the regular structure are also plotted (full lines). The dotted line is the boundary of the light cone. (b)-(e) Detail (on a linear color scale) of the spectral function at the lower edge of band 1, for $\sigma=0.001a$, $\sigma=0.002a$, $\sigma=0.004a$, $\sigma=0.008a$, respectively.}
\end{figure}

Fig. \ref{fig2}(a) is a log-scale intensity plot of the spectral density $A(k,\omega)=-\mbox{Im}(G_y(k,\omega))$ for $\sigma=0.002a$, obtained by diagonalizing Eq. (\ref{eqU2}) on the subspace spanned by bands 1 and 2. The dispersion curves of the two bands for the regular structure are superimposed, as well as the free-space light dispersion. This latter defines the boundary of the light cone: below it, modes of the dielectric slab are fully confined by total internal reflection and ideally lossless. The spectral width along band 1 shows an abrupt increase above the light cone, due to the onset of the intrinsic radiation loss mechanism. A similar abrupt change is visible along band 2 at about $ka/2\pi=0.4$. On the left of that point on the plot, the frequency of band 2 becomes degenerate with points of the same band lying inside the light cone, thus maximizing the mode mixing due to disorder and the subsequent extrinsic radiation losses, as already suggested by Kuramochi et al.\cite{Kuramochi2005} As a consequence, the further increase of the spectral width across the light cone boundary is less abrupt for this band, but still noticeable. Figs. \ref{fig2}(b)-(e) are linear intensity plots of the same quantity in the vicinity of the bottom of the guided band 1, for $\sigma = 0.001a$, $0.002a$, $0.004a$, and $0.008a$, respectively. A feature common to the four cases is the presence of sharp spectral signatures below the bottom of the band of the regular structure. These features are extended along $k$, and extend to increasingly lower frequencies as $\sigma$ is increased. They originate from spatially localized modes induced by disorder, and contribute to form the well known Lifshitz tail in the density of states below the band of the regular crystal.\cite{Halperin1966,Lifshitz1965,Zittartz1966} These features have been experimentally characterized by Le Thomas et al.\cite{LeThomas2009} and the corresponding sharp features in the transmission spectra have been experimentally observed and theoretically studied by modeling the optical response.\cite{Mazoyer2010,Patterson2009}

\begin{figure}[t]
\includegraphics[width=0.5 \textwidth]{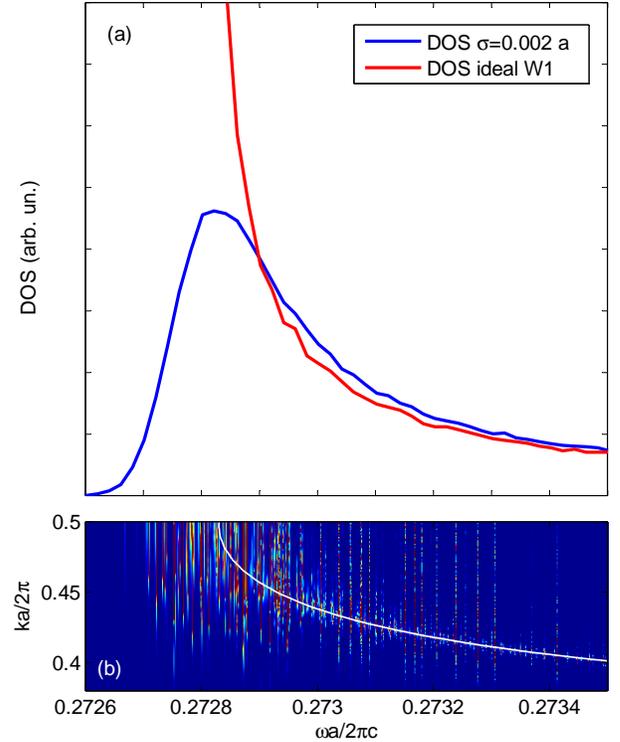}
\caption{\label{fig3} (a) Density of states in the vicinity of the edge of band 1. Blue: $\sigma=0.002a$, averaged over 500 disorder realizations. Red: the corresponding quantity for the regular structure, showing the characteristic Van Hove singularity at the band edge. (b) The spectral function and the band dispersion are plotted as a reference.}
\end{figure}

The density of states can be computed by averaging over several statistical realizations of the disorder configuration. Fig. \ref{fig3}(a) shows the density of states obtained by averaging over 500 realizations, for $\sigma=0.002a$, compared to the density of states of the regular structure (panel (b) shows the spectral density for reference). The Van Hove singularity of the regular crystal is smeared over an interval of the order of $\omega\sigma/(2\pi c)$ and the Lifshitz tail is clearly visible. 

\begin{figure}[t]
\includegraphics[width=0.5 \textwidth]{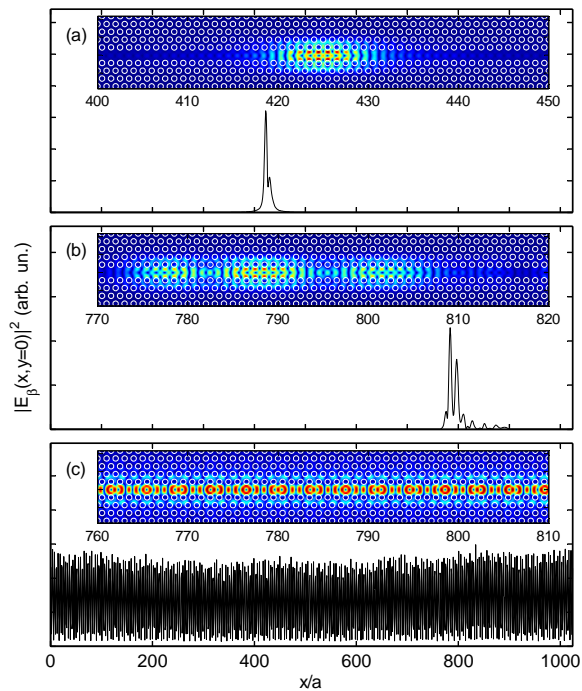}
\caption{\label{fig4} (a)-(c) Electric field spatial profile of three selected eigenmodes computed for $\sigma=0.002a$, at $\omega a/2\pi c =0.27267$, $0.27283$, and $0.27518$, respectively (indicated by horizontal lines in Fig. \ref{fig2}(c)). The main panels display the field intensity computed at the center of each elementary cell, while insets show the full field profile for selected portions of the W1 waveguide.}
\end{figure}

The main innovation brought by the present approach is the possibility to compute the field profile of the actual Maxwell eigenmodes. In connection to the study of disorder, this makes it possible to study the localization properties and brings insight to the question whether Anderson localization of light can occur in PHCs.\cite{Bertolotti2005,John1987,Mazoyer2010,Mookherjea2008,Schwartz2007,Sebbah2006,Topolancik2007,Vlasov1999,Sapienza2010}
We plot the electric field profile of three selected modes of a disorder realization with $\sigma=0.002a$ in Fig. \ref{fig4}(a)-(c) respectively. The three selected frequencies are indicated by dotted lines in Fig. \ref{fig2}(c). The main plot in each panel represents the value of the field intensity at the center of each elementary cell of the guide (in this way the main envelope can be plotted over the full length of the simulated structure), while the insets show the full electric field profile for selected representative regions of the structure. Fig. \ref{fig4}(a) shows the ground eigenmode at $\omega a/2\pi c=0.27267$, which lies well below the band edge and is localized over a few elementary cells of the structure. Fig. \ref{fig4}(b) shows a mode at $\omega a/2\pi c=0.27283$, lying very close to the band edge and displaying a less localized character with few main lobes. Most of the modes computed in this spectral region have a similar character, with several lobes and an overall envelope with exponentially decaying tails, as expected for Anderson localized states. Modes with several lobes have been studied in various kinds of PHCs and are considered as precursors of the extended {\em necklace states}.\cite{Pendry1987,Tartakovskii1987,Bertolotti2005,LeThomas2009,Mazoyer2010,Sebbah2006} Finally, Fig. \ref{fig4}(c) shows a mode at $\omega a/2\pi c=0.27518$, well within the energy range of the guided band, which is delocalized over the whole length of the simulated sample while still showing irregular amplitude fluctuations induced by disorder. This is expected as in the spectral region of this mode, the localization length is much longer than the simulation length $L$ considered here. 

\begin{figure}[t]
\includegraphics[width=0.5 \textwidth]{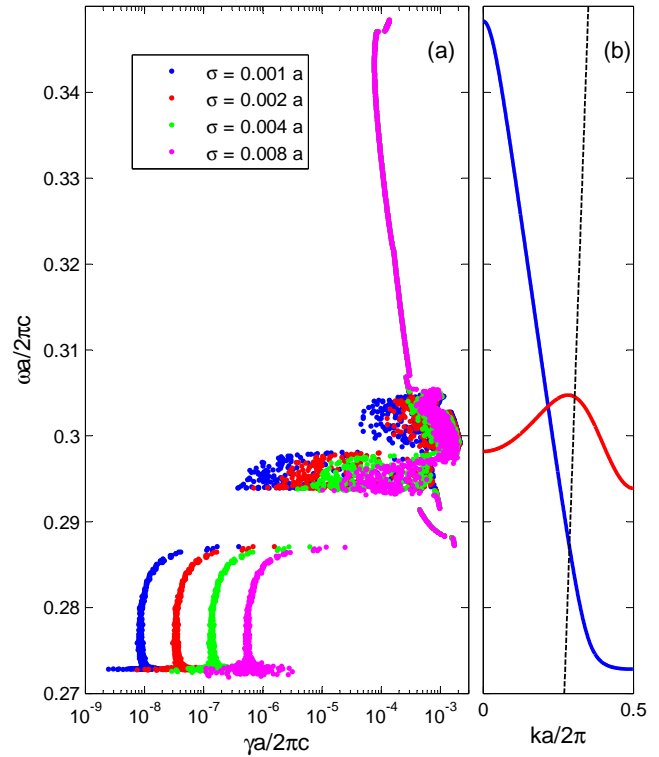}
\caption{\label{fig5} (a) Radiation loss rates of disordered eigenmodes, computed for the four values of $\sigma$ considered in this work. Modes for which the intrinsic radiation loss mechanism is dominant, are those showing a loss rate independent of $\sigma$. (b) Band dispersion (full) and light cone (dotted) are plotted for reference.}
\end{figure}

The radiation loss rate $\gamma_\beta$ for each eigenmode $\beta$ has been computed using Eqs. (\ref{losses}) and (\ref{lossrates}). They are plotted, for one disorder realization, in Fig. \ref{fig5}(a), for $\sigma = 0.001a$, $0.002a$, $0.004a$, and $0.008a$. Panel (b) in this figure shows the band structure and the light line for reference. Let us focus first on frequencies below $0.287$, for which the guided band of the regular structure lies outside the light cone. For this frequency range, radiation losses are extrinsic, as already discussed. For each given value of the disorder amplitude $\sigma$, the rates $\gamma_\beta$ behave quite regularly, except for the lowest frequencies in the Lifshitz tail of the spectrum. There, the values of $\gamma_\beta$ are significantly scattered over about one decade, as one would expect for modes that are strongly localized by the random fluctuations of the underlying crystal structure. A similar trend, with strong variation of the loss rates close to the band edge, was obtained in Ref. \onlinecite{Gerace2004}, although the perturbative approach to disorder adopted in that work cannot predict the fluctuations of $\gamma_\beta$. These large fluctuations are typical of states in the localized part of the spectrum, namely where the localization length is much smaller than the simulation length $L$. Similar fluctuations are observed in the transmission spectrum\cite{Mazoyer2009,Mazoyer2010,Patterson2009} and are more generally expected for all properties of localized eigenmodes, as in the localization lengths discussed below. Above the fluctuating region, the rates $\gamma_\beta$ are approximately constant over a wide frequency range, until they dramatically increase when approaching the boundary of the light cone. The rates $\gamma_\beta$, for the values of the disorder amplitude considered here, are in general very small, as expected for an extrinsic process. When the disorder amplitude $\sigma$ is doubled, the extrinsic rates correspondingly increase by a factor of four. This dependence has already been discussed in the literature.\cite{O'Faolain2007} It can be explained by considering that the overlap integral entering Eq. (\ref{losses}) depends linearly on the disorder matrix element $\langle V\rangle$ to leading order. This can be inferred from Eq. (\ref{lossgm}), if we consider a perturbation expansion of mode $\E_\beta$ where the zeroth order term is a guided Bloch mode, that does not contribute to the overlap integral (\ref{lossgm}) due to orthogonality. For the frequency range above the light line, the eigenmodes of the disordered structure are characterized by much higher loss rates that are now mostly of intrinsic nature and no longer depend on $\sigma$.\footnote{Here, as in Ref. \onlinecite{Gerace2004}, discontinuities appear, due to the singularities in the 1-D density of states entering Eq. (\ref{losses}) for finite $b$. These artifacts would disappear for $b\rightarrow\infty$. In Ref. \onlinecite{Gerace2004} they were smoothed out by averaging the result over a few values of $b$. Here, a different regularization procedure was adopted. The momentum sum in (\ref{losses}) was evaluated as an integral, and $\left|{\cal H}_{\beta\g}^{rad}\right|^2$ was assumed as slowly-varying compared to $\rho_j$. Then, the integral was computed numerically over a momentum grid and $\rho_j$ was integrated analytically within each bin of the grid.} In the frequency range between 0.294 and 0.306, rates of modes related to the spatially-odd band 2 are also present. Here losses show again a partially extrinsic character, especially in the lowest frequency part where a quadratic dependence on $\sigma$ appears. For the discussion below, it is also important to keep in mind that rates in this frequency range are significantly higher than those computed at the bottom of band 1.

\begin{figure}[t]
\includegraphics[width=0.5 \textwidth]{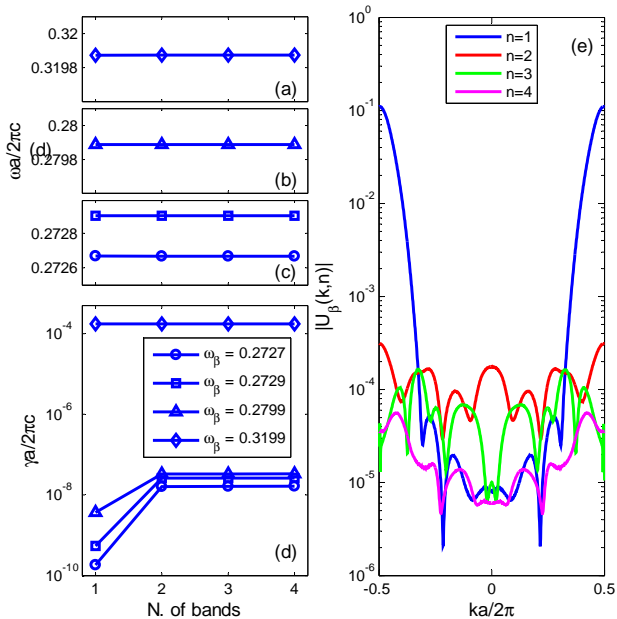}
\caption{\label{fig6} (a)-(c) Eigenfrequencies of four selected eigenmodes computed as a function of the number of bands included in the Bloch-mode expansion, for $\sigma=0.002a$. (d) Computed radiation loss rates of the same modes, as a function of the number of included bands. (e) Coefficients of the Bloch-mode expansion $|U_\beta(k,n)|$ computed for the fundamental mode $\beta=1$ and $n=1,2,3,4$.}
\end{figure}

The origin of the radiation loss rates can be inferred from the nature of the computed eigenmodes. These, according to Eq. (\ref{expansion}) are linear combinations of Bloch modes at different $k$ and different bands. This {\em hybridization} involves all modes of the regular structure, {\em regardless of their frequency spacing $\Delta\omega$} -- the mixing fraction being quantified by $\langle V\rangle(\omega/\Delta\omega)$ to leading order. Fig. \ref{fig6}(a)-(c) shows the convergence of the eigenfrequencies of four selected eigenmodes, at frequencies corresponding to band 1, as a function of the number of bands included in the diagonalization. The convergence is excellent already when only band 1 is included in the Bloch-mode expansion. Convergence of the loss rates, shown in Fig. \ref{fig6}(d), requires instead that the spatially-odd band 2 is also included in the calculation. In particular, for the three lowest frequencies lying in the region of extrinsic radiation losses, the loss rates increase by as much as two decades when band 2 is included. The reason is that the main source of extrinsic losses, close to the edge of band 1, is the hybridization with Bloch modes of the spatially-odd band 2. This mixing would be forbidden by symmetry in the regular structure and is now allowed as disorder breaks the $\hat{\sigma}_{kz}$ invariance. The fourth selected frequency instead corresponds to an eigenmode for which the intrinsic radiation loss mechanism is dominant, and the hybridization with band 2 doesn't thus produce any significant effect. Fig. \ref{fig6}(e) shows the components of the lowest computed eigenmode $\beta=1$, $|U_\beta(k,n)|$. Apart from the dominant contribution of Bloch modes close to $k=\pm0.5$ of band 1, the main contribution comes from band 2 that shows about $3\times10^{-3}$ admixture at the edge of the Brillouin zone. This explains the dominant contribution of band 2 to the extrinsic loss rates, seen in Fig. \ref{fig6}(d). Band 2 losses are several orders of magnitude higher than the extrinsic losses computed when accounting for band 1 only. Therefore, even the slightest admixture of the two bands -- here about 0.3$\%$ -- can completely determine the final loss rate at the band edge. The role of the mixing between spatially-odd and even bands has been already pointed out by Kuramochi et al. \cite{Kuramochi2005} in the framework of perturbation theory, thus restricted to the frequency region where the two modes are degenerate. Our result generalizes the one by Kuramochi et al., by showing that the mixing determines all extrinsic radiation loss rates of the main guided band of a W1 waveguide. The importance of this finding goes well beyond the study of disorder effects. It suggests that radiation losses in all W1-based systems can be reduced by studying designs that minimize the mixing between bands 1 and 2, e.g. by displacing band 2 at higher frequencies. It can indeed be argued that the same mechanism limits the Q-factor of cavities based on the gentle confinement of the W1 guided mode.\cite{Song2005,Felici2010,Kuramochi2006} Our conclusion thus shows a new direction to take for the optimization of such cavity designs.

\begin{figure}[t]
\includegraphics[width=0.5 \textwidth]{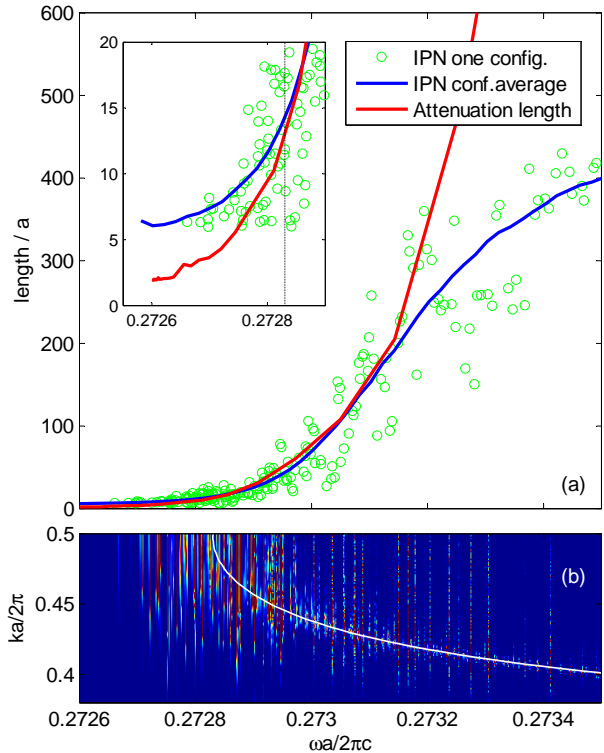}
\caption{\label{fig7} (a) Circles: IPN values of the lowest eigenmodes computed for one disorder realization with $\sigma=0.002a$. Blue line: IPN averaged over 500 disorder realizations. Red line: Average attenuation length of the transmission along the guide. Inset: detail in the vicinity of the edge of band 1 (denoted by a vertical line). (b) The spectral function and the band dispersion are plotted as a reference.}
\end{figure}

To conclude the analysis of the W1 waveguide, we study the localization lengths of the simulated eigenmodes. The localization length in a disordered system can be defined in several different ways.\cite{Kramer1993} One natural definition, when studying eigenstates, is the {\em inverse participation number} (IPN). For a wavefunction $\phi(x)$ in 1-D, the IPN is defined as $I=1/\int|\phi(x)|^4dx$, and has the dimension of a length if $|\phi(x)|^2$ is normalized to 1. The IPN measures the cumulative length over which the field intensity is significantly large. For very singular shapes of the disorder profile, the IPN can differ considerably from the exponential decay length of the tails of $\phi(x)$ -- which is another measure of the localization length. \cite{Kramer1993} For an eigenmode $\HH_\beta(\r)$ of our system, we thus define
\begin{equation}
I_\beta = \left[\int dx|\HH_\beta(x,y=0)|^4\right]^{-1}\,,
\label{ipn}
\end{equation}
where, in order to extract the mode length, we take the integral along the guide core at $y=0$. A second characteristic length related to disorder, typically studied in the context of PHCs,\cite{LeThomas2008,Mazoyer2009,Mazoyer2010,Patterson2009,Vlasov1999,Bertolotti2005} is the exponential attenuation length of the transmission of a monochromatic beam. For this, we introduce the quantity
\begin{equation}
T(x,x^\prime,\omega)=\left|\sum_\beta\frac{\HH^*_\beta(x,y=0)\cdot\HH_\beta(x^\prime,y=0)}{\omega-\omega_\beta+i\gamma_\beta}\right|^2\,,
\label{transmission}
\end{equation}
which is proportional to the transmission of a monochromatic source from $x$ to $x^\prime$. The frequency-dependent exponential attenuation length is defined as the decay constant $L_\omega$ in $T(x,x^\prime,\omega)\propto\exp(-|x^\prime-x|/L_\omega)$. In the numerical calculation of both $L_\omega$ and $I_\beta$, we checked the convergenge of the computed values as a function of the system size $L$, up to the largest value of $L$ considered, in order to detect the onset of finite-size effects. Fig. \ref{fig7}(a) shows the quantities $I_\beta$ and $L_\omega$, as computed for the W1 guide under study. Lines denote configuration averages over 500 disorder realizations, while the circles are IPN values for the eigenmodes of a given realization. The IPN is more affected by the finite size of the simulation and saturates at about $I_\beta=250a$, while the saturation of the attenuation length is not visible on the scale of the plot. This is expected, as the IPN is an extensive property of the eigenmode profile, contrarily to the attenuation length which can be extracted from the slope at a single point. When finite-size effects are negligible, the two quantities roughly coincide. We point out that the attenuation length includes the decay due to radiation losses, while the IPN is an exclusive property of the eigenmodes of Eq. (\ref{eqU2}). In the frequency range of Fig. \ref{fig7}, the computed losses correspond to a negligible contribution to the attenuation length of the transmission, which is thus uniquely determined by the coherent multiple scattering in the propagation along the guide. As for other properties of localized eigenmodes, the IPN displays large fluctuations at each given frequency. This suggests that, independently of the average localization length at a given frequency, rare necklace modes can exist with a profile that extends over the whole sample length $L$.\cite{Pendry1987,Tartakovskii1987,Bertolotti2005,LeThomas2009,Mazoyer2010,Sebbah2006}

Localization in one-dimensional photonic structures has been reported in several experiments.\cite{Bertolotti2005,LeThomas2009,Mazoyer2010,Mookherjea2008,Sebbah2006,Topolancik2007,Sapienza2010} The present result sheds light on the question concerning the nature of light localization in these structures. We have already excluded the extrinsic radiation losses as a cause of attenuation in a W1 waveguide close to the bottom of band 1. It is clear that the origin of localization in our model calculations is to be found in coherent multiple scattering processes, corresponding to the Bloch-mode superposition in expansion (\ref{expansion}). This is of course expected, as in one-dimensional disordered systems localization is always present,\cite{Anderson1985,Kramer1993} and the W1 waveguide shows to be very closely one-dimensional. Indeed, the expansion on a single 1-D Bloch band was sufficient to obtain well converged eigenmodes, while the very small admixture with band 2 was only necessary to account correctly for radiation losses. A relevant question, raised by Vlasov et al.,\cite{Vlasov1999} is whether localization is of the Anderson kind or rather due to a random local energy shift of the band edge induced by large disorder fluctuations. This second mechanism can be pictured as the random alternance in space of pass-band and stop-band regions, and is thus enforced by the Bragg diffraction of the underlying periodic structure, rather than by scattering in an effective medium as in the Anderson localization case. Our approach gives direct access to the spectral function. It shows that, for the disorder amplitudes considered here and corresponding to state-of-the-art samples,\cite{Mazoyer2010,Skorobogatiy2005} the band-edge shift is much smaller than the width of the stop band (see Figs. \ref{fig1} and \ref{fig2}). According to Vlasov et al. then, at least for the disorder model considered here localization should be of the Anderson kind. Comparing the behavior of the IPN and the attenuation length below the band edge, as shown in the inset of Fig. \ref{fig7}(a), supports this conclusion. For decreasing frequency, the attenuation length ceases to follow the average IPN and decreases down to its smallest possible value, essentially given by the unit cell length $a$. For frequencies below this crossover, the attenuation length is no longer determined by the localization length of the eigenmodes, but rather by Bragg diffraction on the PHC, and a similar attenuation length would arise within the stop band also in the case of a regular crystal. The IPN, on the other hand, seems to approach a constant value of a few unit cells, that corresponds to the smallest IPN taken by Anderson localized modes for the given value of $\sigma$, as already observed when directly visualizing the mode in Fig. \ref{fig4}(a).

\section{Application to a L3 coupled-cavity waveguide}
\label{L3}

\begin{figure}[t]
\includegraphics[width=0.5 \textwidth]{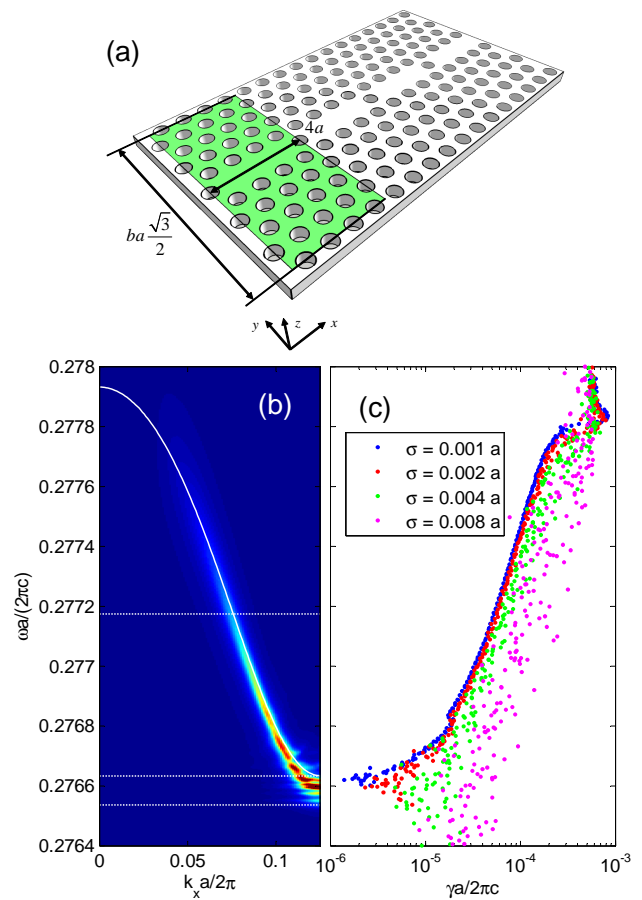}
\caption{\label{fig8} (a) Sketch of the L3 coupled cavity waveguide. The shaded region denotes the supercell used for computing the bands of the regular structure. (b) Spectral function computed for $\sigma=0.002a$, showed as a linear scale color plot. The dispersion of the guided band of the regular structure is plotted as a white line. (c) Radiation loss rates of disordered eigenmodes, computed for the four values of $\sigma$ considered in this work.}
\end{figure}

As an additional test of our method, we consider a chain of L3 coupled cavities, as sketched in Fig. \ref{fig8}(a). The system differs from the W1 waveguide by the presence of one hole every four along the waveguide. This particular kind of {\em coupled-cavity waveguide} has recently been experimentally characterized,\cite{Jagerska2009} in order to determine the origin of out-of-plane radiation losses and the limiting factors to slow light propagation. More generally, coupled-cavity waveguides have been the object of various experimental\cite{Jagerska2009,Jagerska2009a,Notomi2008} and theoretical\cite{Fussell2007,Povinelli2006,Stefanou1998,Yariv1999,Mookherjea2007} studies, because of their potential as effective slow-light channels, thanks to the very flat band expected for an impurity chain.

Here, we start our analysis from the Bloch modes of the regular structure, that we compute as for the W1 waveguide using guided-mode expansion\cite{Andreani2006}, by assuming the supercell sketched in Fig. \ref{fig8}(a). The disorder model and all other parameters are the same as for the W1 waveguide studied above. In particular, the same length $L=1024a$, corresponding to 256 L3 cavities, is assumed. With these parameters, the band of interest\cite{Jagerska2009} goes from $\omega a/2\pi c=0.2766$ to $\omega a/2\pi c=0.2779$, as plotted in Fig. \ref{fig8}(b). Calculations using our method fully converge already when including only this band in the Bloch-mode expansion. This holds also for the radiation loss rates, contrarily to the case of the W1 waveguide, because of the dominant role of the intrinsic radiation loss mechanism as discussed below. The spectral function for $\sigma=0.002a$ is plotted in Fig. \ref{fig8}(b), while Fig. \ref{fig8}(c) displays the radiation loss rates computed for the field eigenmodes, for the disorder amplitude varying from $\sigma=0.001a$ to $\sigma=0.008a$. The Brillouin zone, four times smaller than for a W1 waveguide, now lies completely above the light cone, and radiation losses are mostly determined by the intrinsic mechanism predicted for the regular system.\cite{Fussell2007,Povinelli2006} In particular, the rates vary from a value lower than that of the single L3-cavity,\cite{Andreani2006,Chalcraft2007,Englund2005} at the band bottom, to a higher one at the band top, as predicted in the absence of disorder from the 3-D solution of Maxwell equations\cite{Povinelli2006} or from a tight-binding approach.\cite{Fussell2007}. The dependence on the disorder amplitude $\sigma$ clearly shows that, up to $\sigma=0.004a$, the extrinsic radiation loss mechanism takes over the intrinsic one only in the vicinity of the band bottom. For larger $\sigma$ instead, the extrinsic mechanism becomes relevant over the whole band, and large fluctuations of the rates appear at each given frequency.

\begin{figure}[t]
\includegraphics[width=0.5 \textwidth]{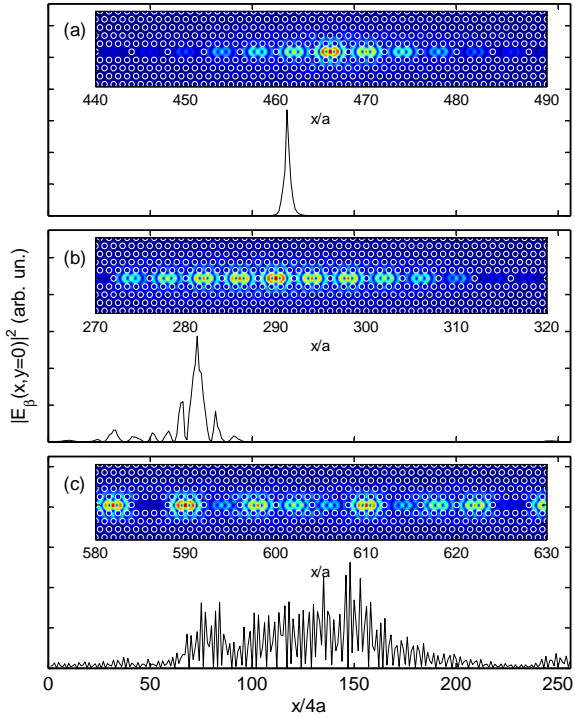}
\caption{\label{fig9}  (a)-(c) Electric field spatial profile of three selected eigenmodes computed for $\sigma=0.002a$. The three corresponding values of $\omega a/2\pi c$ are denoted by horizontal lines in Fig. \ref{fig8}(b). The main panels display the field intensity computed at the center of each elementary cell, while insets show the full field profile for selected portions of the L3 coupled-cavity waveguide.}
\end{figure}

Fig. \ref{fig9} shows the electric field intensity computed for three selected eigenmodes, denoted by horizontal lines in Fig. \ref{fig8}(b). The ground eigenmode in Fig. \ref{fig9}(a) is localized over the length of a few L3-cavities. Modes at higher frequencies displayed in panels (b) and (c) are progressively more delocalized, again as expected for a one-dimensional disordered system. Notice that each mode has the shape of a slowly varying envelope function modulating the field profile of the ground mode of a L3 cavity,\cite{Chalcraft2007} as in tight-binding calculations.\cite{Fussell2007}

\begin{figure}[t]
\includegraphics[width=0.5 \textwidth]{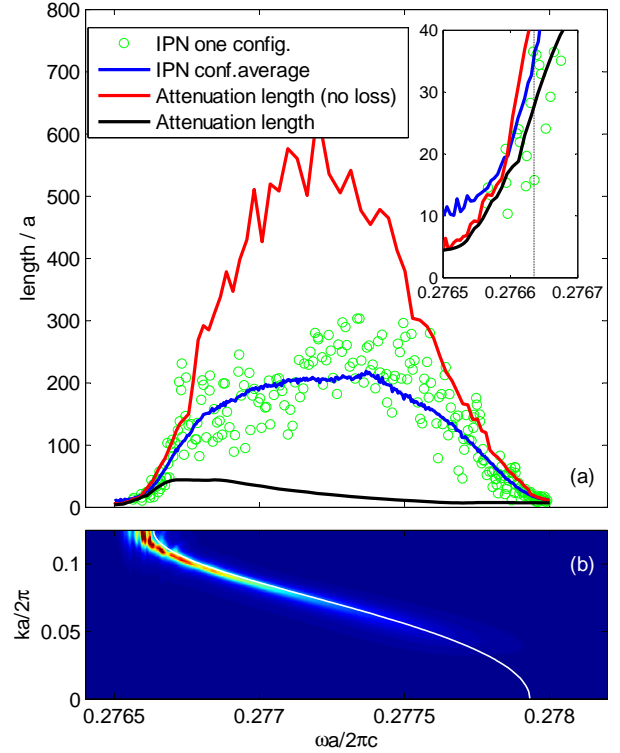}
\caption{\label{fig10}  (a) Circles: IPN values of the eigenmodes computed for one disorder realization with $\sigma=0.002a$. Blue line: IPN averaged over 500 disorder realizations. Red line: Average attenuation length of the transmission along the guide. Inset: detail in the vicinity of the lower band edge (denoted by a vertical line). (b) The spectral function and the band dispersion are plotted as a reference.}
\end{figure}

As for the W1 waveguide studied in the previous Section, the modes displayed in Fig. \ref{fig9} are computed without accounting for losses [which are computed at a later stage through Eq. (\ref{losses})]. Contrarily to the W1 waveguide however, here the attenuation length corresponding to the computed radiation loss rate is actually shorter than the localization length of the modes, when considering the least localized modes at the band center. Hence, for these modes the result of our Bloch-mode expansion is not fully consistent with the initial assumption of lossless Bloch modes. To better understand this issue, we show in Fig. \ref{fig10}(a) the localization lengths computed from the IPN and the attenuation length computed from the decay of the transmission (\ref{transmission}). Fig. \ref{fig10}(b) shows the spectral function for reference. The attenuation length is evaluated both by accounting for the computed loss rates $\gamma_\beta$ (black curve) and by assuming a vanishing loss rate for all modes (red curve). This latter coincides with the average IPN over a wide range of frequencies close to the band extrema, while at the band center the IPN starts being affected by the finite length of the simulated structure, similarly to the case of the W1 waveguide. The attenuation length including losses is instead dramatically suppressed, except for a narrow region at the band bottom (see inset). We argue that, in the region where the two curves differ significantly, the localized eigenmodes computed here are actually ill-defined. Radiation loss rates of Bloch modes are so high that they should be accounted for already in the multiple scattering process from which the eigenmodes of the disordered system arise. This could be achieved by replacing the guided-mode expansion with a full expansion on guided plus radiative modes of the slab, or simply with an expansion on 3-D plane waves. The inset of Fig. \ref{fig10}(a) shows a detail of the region below the band edge, displaying the same trend already observed for a W1 waveguide in Fig. \ref{fig7}(a), although now the minimal IPN value is larger than in the W1 case, as expected from a longer unit cell.

\begin{figure}[t]
\includegraphics[width=0.5 \textwidth]{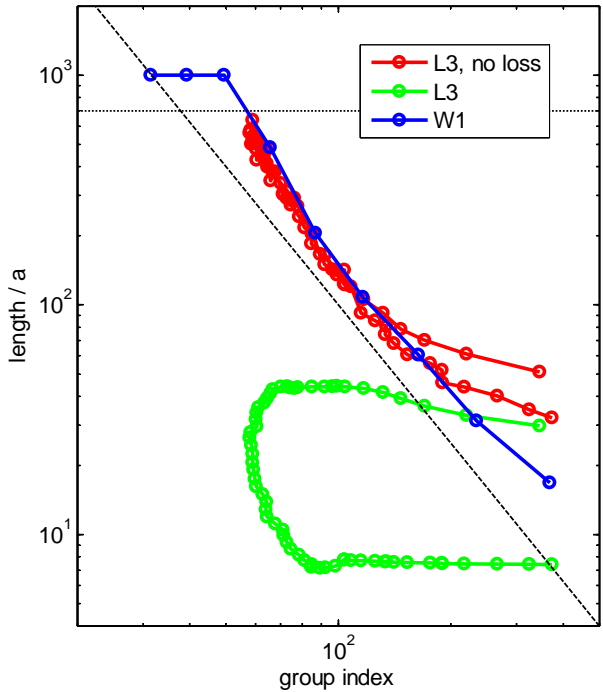}
\caption{\label{fig11} Double logarithmic plot of the attenuation length of the transmission as a function of the nominal group index. The horizontal line indicates the limit set by the finite simulation length, while the diagonal line is proportional to $n_g^{-2}$ and is plotted as a guide to the eye. Blue: W1 waveguide. Red: L3 coupled-cavity waveguide neglecting radiation losses. Green: L3 coupled-cavity waveguide accounting for radiation losses.}
\end{figure}

We conclude by making some considerations on the influence of disorder on slow-light propagation, both in the W1 and in the coupled-cavity waveguide. Slow-light has been one of the main driving forces of the development of one-dimensional PHC structures.\cite{Baba2008,Krauss2007,Notomi2001,Vlasov2005} Ideally, close enough to the bottom of the guided band, the group velocity of a narrow light pulse, defined as $v_g=d\omega_k/dk$, can become arbitrarily small. Disorder hinders this ideal behavior by producing a strong attenuation of the transmitted signal, as discussed above. Past experimental studies\cite{Engelen2008,Kuramochi2005,LeThomas2009,Patterson2009} have focused on the dependence of the attenuation length on the group index, defined as $n_g=c/v_g$. All recent theoretical studies\cite{Hughes2005,Mazoyer2009,Mazoyer2010,Patterson2009,Wang2008,Povinelli2004} show that the dominant functional dependence of the attenuation length close to the band bottom is on $n_g^{-2}$. This behaviour is induced by disorder through the mechanism of coherent backscattering, determined by second and higher-order coherent scattering processes, which is a precursor of localization.\cite{Kramer1993} The $n_g^{-2}$-dependence was confirmed in several experiments.\cite{Engelen2008,Kuramochi2005,O'Faolain2007} It was also pointed out\cite{Mazoyer2009,Mazoyer2010,Wang2008} that, very close to the band bottom, the onset of localization makes multiple scattering beyond second order important and the $n_g^{-2}$-dependence consequently breaks down for high $n_g$, with the attenuation length that levels off to its minimal value given by the lattice parameter $a$. In Fig. \ref{fig11}, we plot the computed attenuation length $L_\omega$ for $\sigma=0.002a$ as a function of the {\em nominal} group index $n_g$, both for the W1 and for the coupled cavity waveguides. By ``nominal'', it is meant the group index that would result at frequency $\omega$ from the dispersion of the ideal band. The plot thus does not include frequencies below the band bottom. The result for the W1 waveguide (blue) reproduces the result obtained by Mazoyer et al.\cite{Mazoyer2009,Mazoyer2010} For the coupled cavities, we display two curves corresponding to the case including radiation losses (green), and to the unphysical but illustrative lossless situation (red) respectively. This latter coincides with the W1 result, except for the largest values of $n_g$ considered, that arise close to the band bottom. The interpretation of this result is straightforward. A small group index corresponds to a frequency well above the band bottom, where eigenmodes have a localization length much longer than the elementary cell. In this case, independently of the length of the elementary cell, the envelope function of the eigenmodes will be localized according to the John mechanism,\cite{John1987} with the localization length solely determined by the curvature of the Bloch-mode dispersion, thus by the group index. When approaching the band bottom instead, the localization length becomes smaller and eventually reaches its lower bound corresponding to a small number of elementary cells, as shown from the analysis of the IPN. In this limit, the localization length of the coupled-L3 system is four times longer than that of the W1 waveguide, at corresponding $n_g$. This limiting behavior can be inferred by comparing the upper branch of the red curve (corresponding to the band bottom) and the blue curve in Fig. \ref{fig11}. When accounting for intrinsic radiation losses in the L3-coupled system, the attenuation length is dramatically reduced except for the region close to the band bottom, where losses are the smallest and disorder concurs significantly in determining the attenuation. Following Mazoyer et al.,\cite{Mazoyer2009,Mazoyer2010}, a diagram of the kind displayed in Fig. \ref{fig11} sets the maximum length scale over which light propagation can still be ballistic, and thus the slowdown predicted for the ideal system can still occur. Then, from this analysis we can conclude that, in spite of the increased loss rates, a L3-coupled waveguide can still bring an advantage over the W1 waveguide when close to the band bottom, by increasing four times this maximum length for a given (large) group index.

\section{Conclusions}
\label{conclusions}

The main result of the present analysis is the demonstration that Boch-mode expansion can be an extremely effective and reliable tool for studying the properties of the eigenmodes of Maxwell equations, in PHCs where disorder or small systematic variations of the otherwise periodic structure are present.

The approach relies on a method to compute the Bloch modes of the regular PHC. Ideally, this first elementary step should be accomplished by exact diagonalization of the Maxwell problem within a single elementary cell, for example by using a finite element approach. For the scope of the present work however, we preferred to adopt the guided-mode expansion method,\cite{Andreani2006} because of its reliability at a minimal computational cost. In this case, however, out-of-plane propagating modes are not included in the expansion, and the radiation losses of the final Maxwell eigenmodes must be computed non self-consistently within a perturbative approximation at a later stage.

The general advantage of Bloch-mode expansion lies in the optimal choice of the Bloch-mode basis, for situations where the variations from the periodic structure are small. Then, an expansion restricted to a very limited number of spectrally close Bloch bands, can produce excellent convergence, thus practically reproducing the exact solution of Maxwell equations at a reasonable computational cost. 

As an example, we have applied the method to study the effect of disorder in a W1 line-defect waveguide and a L3 coupled-cavity waveguide. The method gives direct access to the shape and spectrum of the eigenmodes, from which the localization properties, and the mechanism underlying radiation losses have been characterized. 

One very important result of this work -- apart from the assessment of the Bloch-mode expansion method -- is the explanation of the origin of radiation losses of the spatially even guided modes of a W1 waveguide close to the band bottom. Radiation losses are mainly determined by the hybridization of these modes with those of the guided band of opposite spatial parity. A very small admixture of the two bands is sufficient to increase the radiation loss rate close to the band edge by two decades. Hence, the effect is dominant in spite of the frequency separation of the two bands that hybridize -- a feature overlooked by previous analyses. This result was only possible thanks to the separate analysis of the contribution of different Bloch bands. It unveils the physical mechanism limiting the quality factor of all structures based on a W1 waveguide. In particular, all designs of high-Q cavities based on slow modulation of a W1 waveguide should be affected. We suggest that a significant increase in the quality factor of such structures can be obtained by studying a waveguide design that reduces the hybridization, e.g. by maximizing the frequency separation between the two guided bands. 

The Bloch-mode expansion has a very broad spectrum of potential applications. These range from the realistic modeling of high-Q cavities, to the study of radiation-matter coupling and cavity quantum electrodynamics effects for systems of several quantum dots coupled to the modes of PHC cavities or waveguides. All these application should take advantage of the possibility of computing the actual eigenmodes of Maxwell equations, thus making the Bloch-mode expansion an election method for the modeling and design of photonic systems.

\begin{acknowledgments}
I am very grateful to Romuald Houdr\'e, Nicolas Le Thomas, Jana J\'agersk\'a and Marco Felici for several enlightening discussions. This work is supported by NCCR Quantum Photonics (NCCR QP), research instrument of the Swiss National Science Foundation (SNSF).
\end{acknowledgments}

\appendix

\section{Guided-mode expansion}

The modes $\E_{\k n}(\r)$ of the regular PHC, and their eigenfrequencies $\omega_{\k n}$ can be computed using the guided-mode expansion method introduced by Andreani and Gerace \cite{Andreani2006}. The modes are then expanded on the basis of the guided modes of an effective homogeneous dielectric slab, characterized by dielectric constants $\bar\epsilon_j$, with $j=1,2,3$ indicating the lower cladding, the PHC layer and the upper cladding respectively. The quantities $\bar\epsilon_j$ are the spatial averages of the actual dielectric profiles in each layer. Guided modes $\E^{(g)}_{\k+\G,\alpha}(\r)$ are labeled by their total in-plane momentum $\g=\k+\G$ and mode index $\alpha=1,2,\ldots$. Then, the PHC modes are expanded as
\begin{equation}
	\E_{\k n}(\r)=\sum_{\G,\alpha}c_n(\k+\G,\alpha)\E^{(g)}_{\k+\G,\alpha}(\r)\,.
\label{gmexpansion}
\end{equation}
In this work we give explicit expressions only for guided modes having TE polarization, as we assume quasi-TE PHC modes in the results that we present in Sections \ref{application} and \ref{L3}. More general expressions, for modes where TE and TM polarizations are mixed, can be obtained in a straightforward way by following the guidelines in Ref. \onlinecite{Andreani2006}.

For the calculation of the matrix elements $V_{\k n,\k^\prime n^\prime}$, only the expression of $\E_{\k n}(\r)$ for $\r$ in the slab is needed, as we assumed $\delta\epsilon(\r)=0$ in the claddings. By further assuming a structure symmetric with respect to a mirror reflection $\hat{\sigma}_{xy}$, namely equal upper and lower claddings, the guided-mode expansion reads
\begin{equation}
	\E_{\k n}(\r)=\sum_{\mu}c_n(\mu)i\frac{\omega_\mu}{c}\bm{\epsilon}_\g\frac{e^{i\g\cdot\bm{\rho}}}{\sqrt{S}}2\mbox{Re}(A_{2\mu})\cos(q_\mu z) \,,
\label{gmexpansion2}
\end{equation}
where $\mu=(\k+\G,\alpha)$ is the global index for the guided mode, $\bm{\rho}$ is the in-plane position vector and $z$ the coordinate orthogonal to the plane, $q_\mu=\sqrt{\bar{\epsilon}_2\omega_\mu^2/c^2-g^2}$, and $S$ is a normalization area that will disappear from the final result. The complex coefficient $A_{2\mu}$ is defined in Ref. \onlinecite{Andreani2006}. We have introduced the unit polarization vector $\bm{\epsilon}_\g=\mathbf{z}\times\g$ of the guided mode, where $\mathbf{z}$ is the unit vector along the $z$-axis. By replacing (\ref{gmexpansion2}) into (\ref{matel}), we obtain
\begin{equation}
	V_{\k n,\k^\prime n^\prime}=\sum_{\mu,\mu^\prime}c^*_n(\mu)c_{n^\prime}(\mu^\prime)\delta\epsilon(\g^\prime-\g)M_{\mu\mu^\prime} \,,
\label{matel2}
\end{equation}
where
\begin{eqnarray}
M_{\mu\mu^\prime}&=&2\frac{\omega_{\mu}\omega_{\mu^\prime}}{c^2}\bm{\epsilon}_{\g}\cdot\bm{\epsilon}_{\g^\prime}\left[A_{2\mu}^*A_{2\mu^\prime}I_{2-}+\mbox{Re}(A_{2\mu}A_{2\mu^\prime})I_{2+}\right]\,,
\label{Mmu}\\
I_{2\pm}&=&2\frac{\sin\left[(q_\mu\pm q_{\mu^\prime})\frac{d}{2}\right]}{q_\mu\pm q_{\mu^\prime}}\,.
\label{I2}
\end{eqnarray}
In Eq. (\ref{matel2}) enters the Fourier transform of the dielectric perturbation, defined as
\begin{equation}
\delta\epsilon(\g)=\frac{1}{S}\int_Sd\bm{\rho} e^{i\g\cdot\bm{\rho}}\delta\epsilon(\bm{\rho},z=0)\,.
\label{FTepsilon}
\end{equation}
Here, we make use of the assumption, introduced in Section \ref{application}, that the disorder profile $\delta\epsilon(\r)$ is constant along $z$ within the slab. Within the disorder model introduced in that Section, circular holes of varying radius and position are assumed, and the Fourier transform (\ref{FTepsilon}) can be evaluated analytically. We start from the Fourier transform of the disordered dielectric profile $\epsilon^\prime(\bm{\rho},0)$
\begin{eqnarray}
\epsilon^\prime(\g)&=&\frac{1}{S}\int_Sd\bm{\rho} e^{i\g\cdot\bm{\rho}}\epsilon^\prime(\bm{\rho},0)\nonumber\\
&=&\bar{\epsilon}_2\delta_{\g,0}+(1-\bar{\epsilon}_2)\sum_m\frac{1}{S}e^{i\g\cdot\bm{\rho}_m}\frac{2\pi r_m}{g}J_1(gr_m)\,,
\label{FTepsilonprime}
\end{eqnarray}
where $r_m$ and $\bm{\rho}_m=(x_m,y_m)$ are respectively the radius and the in-plane position of the $m$-th hole of the structure, $J_1$ is the Bessel function of the first kind of order 1, and the sum runs over all holes present in the PHC. An analogous expression holds for the Fourier transform of the regular dielectric profile $\epsilon(\g)$, if $r_m$ and $\bm{\rho}_m$ are replaced by $r$ and the positions $\bm{\rho}^{(0)}_m=(x^{(0)}_m,y^{(0)}_m)$ of the holes in the regular structure. Then, Eq. (\ref{FTepsilon}) is simply expressed as
\begin{equation}
\delta\epsilon(\g)=\epsilon^\prime(\g)-\epsilon(\g)\,.
\label{FTepsilon2}
\end{equation}
We point out that the present method can be at least formally extended to more elaborate disorder models, with holes of non-circular shape and possibly with a $z$-dependence of the dielectric profile (e.g. tapered holes or roughness along $z$). For this, it is enough to account for the corresponding $\delta\epsilon(\r)$ in the integral (\ref{matel}). For an arbitrary disorder shape however, this integral must in general be computed in a fully numerical fashion, which might make the method computationally very demanding. 

In order to compute the radiation loss rate of each mode, we generalize the expression introduced by Andreani et al. \cite{Andreani2006}. In analogy to the Fermi golden rule in quantum mechanics, here the loss rate is determined by its overlap matrix element with leaky modes of the effective slab, through Eq. (\ref{losses}). Within the guided-mode expansion, the matrix element ${\cal H}_{\beta\g}^{rad}$ reads
\begin{equation}
{\cal H}_{\beta\g}^{rad}=\frac{\omega_\beta^2}{c^2}\int d\r\epsilon(\r)\E^*_\beta(\r)\cdot\E^{rad}_{\g\lambda j}(\r)\,,
\label{lossgm}
\end{equation}
where $\E^{rad}_{\g\lambda j}(\r)$ are the leaky modes of the effective slab, described in detail in Ref. \onlinecite{Andreani2006}, $\lambda$ runs over the two polarizations TE and TM, and $j=1,3$ runs over the lower and upper cladding layer. By using the two expansions (\ref{expansion}) and (\ref{gmexpansion}), we can express the matrix element as
\begin{equation}
{\cal H}_{\beta\g}^{rad}=\sum_{\k^\prime,n}\sum_{\G^\prime,\alpha}U_\beta^*(\k^\prime, n)c_n^*(\k^\prime+\G^\prime,\alpha){\cal H}_{\g^\prime\g}^{rad}\,,
\label{lossgm2}
\end{equation}
where
\begin{equation}
{\cal H}_{\g^\prime\g}^{rad}=\frac{\omega_\beta^2}{c^2}\int d\r\epsilon(\r)\E^{(g)*}_{\g^\prime\alpha}(\r)\cdot\E^{rad}_{\g\lambda j}(\r)\,.
\label{lossgm3}
\end{equation}
Because of the translational symmetry of the lattice, the above integral is proportional to $\delta_{\k\k^\prime}$, thus lifting the sum over $\k^\prime$ in (\ref{lossgm2}). Details for the evaluation of Eq. (\ref{lossgm3}) are given in Ref. \onlinecite{Andreani2006}.


\end{document}